# A Survey on BGP Issues and Solutions

Amit Narayanan
Applied Computer Science
The University of Winnipeg
515 Portage Avenue
Winnipeg, MB R3B 2E9
anarayan@uwinnipeg.ca

*Abstract*—BGP is the *de facto* protocol used for inter-autonomous system routing in the Internet. Generally speaking, BGP has been proven to be secure, efficient, scalable, and robust. However, with the rapid evolving of the Internet in the past few decades, there are increasing concerns about BGS's ability to meet the needs of the Internet routing. There are two major limitations of BGP which are its failure to address several key security issues, and some operational related problems. The design and ubiquity of BGP have complicated past efforts at securing inter-domain routing. This paper surveys the past work related to BGP security and operational issues. We explore the limitations and advantages of proposed solutions in these two limitations.

## I. INTRODUCTION

The Internet is a global, decentralized network comprised of many smaller interconnected networks. Networks are largely comprised of end systems, referred to as hosts, and intermediate systems, called routers. Information travels through a network on one of many paths, which are selected through a routing process. Routing protocols communicate reachability information (how to locate other hosts and routers) and ultimately perform path selection.

The Border Gateway Protocol (BGP) is an inter-autonomous system (AS) routing protocol. An autonomous system is an administrative domain. That is, it is a network or group of networks under a common administration and with common routing policies. BGP is used to exchange routing information in the Internet and is the protocol used by default to communicate between Internet service providers (ISP). Customer networks, such as universities and corporations, usually employ protocols known as Interior Gateway Protocol (IGP) to exchange routing information within their networks. Examples of IGPs are Routing Information Protocol (RIP) and Open shortest Path Protocol (OSPF). Customers connect to ISPs, and ISPs use BGP to exchange customer and ISP routes. A network under the administrative control of a single organization is called an autonomous system (AS) [18]. There are two types of routing, intra-domain routing which is the process of routing within an AS, and inter-domain routing which is the process of routing among different ASes. BGP is the dominant inter-domain routing protocol on the Internet (BGP) [45]. BGP has been deployed since the commercialization of the Internet, and version 4 of the protocol has been in wide use for over a decade. There are two variations of BGP: Interior BGP (IBGP), which is used by ISPs to exchange routing information within an AS; and External BGP (EBGP), which is used to exchange routes among autonomous systems. Figure 1 illustrates the difference IBGP and EBGP.

BGP is a simple protocol and it generally works well in practice. Thus, it has played a fundamental role within the global Internet [17], despite providing no performance or security guarantees.

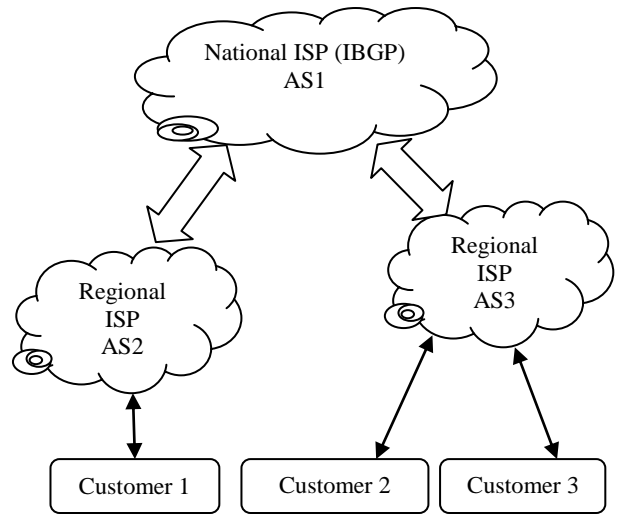

Figure 1: IBGP vs. EBGP

Unfortunately, due to the limited guarantees provided by BGP, it sometimes causes serious instability and outages. Unlike other routing protocols that have limited failing impact and scope, BGP problems may result in significant and widespread damage.

Current research on BGP focuses on addressing and resolving issues related with both operational and security. Operational concerns relating to BGP, such as scalability (i.e when routing tables grow very huge), convergence delay (i.e., the time required for all routers to have a consistent view of the network), routing stability and oscillation, and performance, have been addressed the most and were the major concern for both research and industry communities.

On the other hand, much of the occurring security research has focused on the issues related to authentication, authorization, integrity, confidentiality, and validation of BGP messages. These two fields of operational issues and security research are inherently connected. That is, successes and failures in each domain affect the other domain also and resolving an issue related to one domain is helpful to both communities.

This paper investigates ongoing research in inter-domain routing from the aspects of operational practice, standards activity, and security, exposing the similarities and differences in the proposed approaches towards building a more efficient and secure Internet infrastructure.

The rest of the paper is organized as follows. The next section presents background about BGP from operational and security points of view. Section III describes BGP security issues and proposed solutions. In Section IV we focus on issues related to BGP functionality that have been addressed in the literature. Finally, Section V concludes the paper.

## II. BACKGROUND

The Internet is a global, decentralized network comprised of tens of thousands of smaller interconnected networks. These networks are known as Autonomous Systems (ASes). The Border Gateway Protocol (BGP) is the routing protocol used to exchange information between these ASes. Each BGP-speaking router sends an announcement message when a new route is discovered, and a withdrawal message is also sent when a route no longer exists. BGP is also a path-vector protocol. That's is, when a router advertises a path, it adds its AS number to the beginning of the AS path. The BGP is also policy-based; each router selects the best possible BGP route for each destination prefix and may apply complex policies for selecting such a route. It also decides whether to advertise the route to a neighboring router in another AS.

In this section, we present an overview of the issues related to the inter-domain routing in the Internet and describe some of the BGP's major problems. These problems mainly caused by the following reasons: (i) uncertainty about the relationship between IP prefixes and the AS numbers of the ASes who manage them, (ii) the use of the Transmission Control Protocol (TCP) as the underlying transport protocol, and (iii) the potential to tamper with route announcements in order to subvert BGP routing policy.

### A. IBGP scalability issue

An autonomous system that deploys Internal BGP (IBGP) must have all of its routers that speak iBGP connected to each other through IBGP sessions in a full mesh so that each router can communicate directly with others. Since the full-mesh configuration requires that each router maintain a session to every other router in the network, the number of sessions is $O(n^2)$ where $n$ is the number of routers that speak IBGP. When the network grows and number of routers increases, the number of sessions may degrade the performance of routers, due either to inefficient resources such as memory, or very high CPU utilization.

To overcome this issue, two solutions were proposed route reflectors and confederations. Both techniques reduce the number of IBGP sessions need to be maintained in the network and consequently reduce processing overhead. While route reflectors are considered a pure performance-enhancing technique [17], route confederations are mainly used to implement more fine-grained policy.

However, these alternatives can introduce a set of problems of their own, including the following:
1. route oscillation;
2. sub-optimal routing; and
3. increase of BGP convergence time [9]

### B. Instability issue

Since routing table have to be consistent with network, the routing tables managed by a BGP implementation are adjusted continually to reflect actual changes in the network infrastructure. Examples of such changes are links breaking and being restored or routers going down and coming back up. These events happen almost continuously in the network as a whole and they are considered normal. However, the frequency of these events should be low for a specific router or link. When a router is misconfigured or mismanaged then it may get into a frequent cycle of going down (withdrawal) and then up (reannouncement). Consequently, this pattern of repeated route withdrawal and then reannouncement can result in abnormal activity in all the routers that know about the broken link, as the same route is continuously injected and withdrawn from the routing tables. This problem is known as *route flapping*.

### C. Routing table growth issue

One of the key issues faced by BGP is the growth of the routing table. This issue comes into picture when the routing table grows to the point where some older, less capable, routers cannot cope with the resource requirements for maintaining the routing table. Thus, these routers will cease to be effective gateways between the parts of the Internet they connect. Furthermore, larger routing tables usually take longer time to stabilize on a path when a major routing table change occurs, which affects the network service reliability and availability.

### D. Load-balancing issue

Another factor causing this growth of the routing table is the need for load balancing of multi-homed networks. It is not a trivial task to balance the inbound traffic to a multi-homed network across its multiple inbound paths, due to limitation of the BGP route selection process. For a multi-homed network, if it announces the same network blocks across all of its BGP peers, the result may be that one or several of its inbound links become congested while the other links remain under-utilized, because external networks all picked that set of congested paths as optimal. Like most other routing protocols, the BGP protocol does not detect congestion.

*E. Security Issues*

In the last decade, several major incidents and attacks have been reported regarding compromising of the routing infrastructure on the Internet. Many of these incidents and attacks have resulted in issues such as misrouted traffic and denial of services (DoS). One of the subjects that have been studied widely is the prefix hijack attack in which hackers update BGP routing tables with false origin information which causes serious consequences when this information are propagated. These attacks need to be detected early and accurately so that their propagation through the Internet can be stopped and damage can be mitigated quickly. Early approaches to develop BGP security extensions have failed, but new research directions in heuristic, data driven approaches to suppressing erroneous and malicious BGP messages show some practical promise.

### III. BGP SECURITY ISSUES

The BGP security issues have been widely investigated by the research community. The Internet Engineering Task force (IETF) has discussed some of the main security problems related to BGP, proposed possible In attempt to overcome BGP security issues, several extensions for BGP have been proposed. Kent et al. [19] proposed a secure, scalable, deployable architecture (S-BGP) for an authorization and authentication system that addresses most of the security problems associated with BGP. They discussed the vulnerabilities and security requirements associated with BGP, described the S-BGP countermeasures. They also provided a comparison of their architecture to other approaches that have been proposed, analyzed the performance implications of the proposed countermeasures, and addressed operational issues.

White et al. [9] proposed secure origin BGP (soBGP), where origin authentication is accomplished in an oligarchy PKI similar to that in S-BGP. The main difference between S-BGP and soBGP is that soBGP does not use cryptographic mechanisms to secure the authenticity of the entire AS PATH. It instead verifies AS paths against a database of AS-to-AS routing relationships. soBGP validates the correctness and authorization of the data carried within BGP, and also prevents the sorts of attacks resulting from misconfiguration or intentional insertion of bad data into the Internet routing system.

Kruegel et al. [7] proposed a method for detecting malicious inter-domain routing update messages by monitoring BGP traffic.

Goodell et al. propose IRV [5] that works with BGP to maintain dedicated verification servers and to verify the authenticity of BGP advertisements.

Yu et al. propose a mutual trust-based scheme to evaluate authenticity of BGP advertisements [10]. Their method is incrementally deployable, protects against shilling attacks, and deters malicious operator behavior.

Aiello et al. also address the problem of origin authentication through the use of Address Delegation Graph (ADG) [4]. Subramanian et al. propose a method called Listen and Whisper [8], which eliminates a large number of problems due to router misconfigurations and can detect and contain isolated adversaries that propagate even a few invalid route announcements.

Hu et al. propose a new protocol called Secure Path Vector (SPV) [6] focus on securing BGP update messages against attacks and addresses AS PATH authentication through the use of one-time signatures and symmetric cryptographic primitives. They also limit the use of expensive public-key cryptography.

### IV. BGP OPERATIONAL ISSUES

EBGP is a path vector protocol, which means that loops in routing paths are detected and avoided by checking for multiple occurrences of an AS in the AS_PATH at each BGP node.

However, this scheme cannot be used to detect loops in IBGP since all the speakers belong to the same AS. Thus, to avoid loops in IBGP, every BGP router is required to maintain an IBGP session with all other BGP routers within the AS. Obviously maintaining a full mesh of IBGP sessions is not scalable. To overcome this scalability issue, there are two common IBGP configuration schemes: *AS confederations* and *route reflections* [17]. Using route reflections and AS confederations may cause several operational problems such as routing oscillations.

Varadhan et.al. [20] were the first to discuss the possibility of persistent route oscillations in BGP even in simple topologies. They showed that the oscillation cause was not the policy configuration of one AS alone; but it occurs due to interaction between the policies of several ASes. They further showed that these anomalies can occur without any misconfiguration and they are difficult to diagnose and resolve.

Griffin et.al. [11] introduced the Stable Paths Problem (SPP) as a formal model for vector routing model in general and for BGP specifically. They used their framework to provide a sufficient condition for protocol convergence, which is the absence of dispute wheels. Unfortunately, they showed that the problem of detecting whether stable routing exists, given all the policies in the network, is NP-complete. They also showed that the existence of a stable solution does not automatically imply that a routing protocol can find it. Later Griffin et al. [13] showed that route oscillations can occur even without taking MED.

There have been several follow ups to investigate these routing anomalies. One of the approaches to eliminate

MED oscillations, was proposed by Basu et al. [1]. They proposed to change the protocol such that the oscillation problem vanishes. Basu et al. [1] also presented a counterexample for the solution provided by Walton et al. [12]. However, Griffin [37] showed that the method proposed in [1] has had scaling issues. In [37], Griffin et al. analyzed the oscillations and loops due to path asymmetry using a graph theoretic approach. They further proved that detecting such anomalies is NP hard.

In [14], Musunuri et al. proposed to modify the IBGP protocol to eliminate these anomalies. Their method is based on applying IBGP with some restrictions on the IBGP configuration. They also assume a full mesh of IBGP sessions among all the border speakers. However, this is not very practical as it seems similar to assuming a full mesh of IBGP sessions between all the BGP speakers.

In [16], Gobjuka studied forwarding loops caused by IBGP misconfiguration. He showed that finding forwarding loops in IBGP networks is inherently hard. He further proposed a polynomial-time algorithm for clustering ASes and he showed that the AS configured using his method results in forwarding-loop free network.

Later, in [15] Musunuri et al. proposed another changes to IBGP to solve the problems due to both MED attribute and path asymmetry.

Gao and Rexford [3] studied the Internet economics and showed that it could naturally guarantee route stability. Specifically, they show that a hierarchical business structure underlying the graph representation of the AS, in combination with routing policies is sufficient for protocol convergence. In this structure, they followed the customer-provider relationships between different ASes.

In a follow-up research [2], Feamster et.al. improved this result and demonstrate that certain rankings that are commonly used in practice may not ensure routing stability. Further, they proved that the routing system will converge to a stable path when providers can set rankings and filters autonomously.

## V. CONCLUSION

BGP has been very successful in providing stable and robust inter-domain routing. BGP is widely deployed globally and it is the only Inter-domain routing protocol in wide use, consequently, it has gained increasing interest in both research and industrial communities.

In this article, we first provide some background about BGP and related operational and security issues. Then we investigated some of the work that has been done to address these concerns.